\documentclass[11pt]{article}  
\usepackage{menuproc}
\usepackage{cite}
\usepackage{epsfig}
\usepackage{amsmath,amssymb}
\begin{document}

\titlematter{Application of Roy's equations to analysis of $\pi\pi$
experimental data}
{B. Loiseau$^a$, R. Kami\'nski$^b$ and L. Le\'sniak$^b$}
{$^a$ LPNHE, Univ. P. \& M. Curie,  Paris, France\\
$^b$ Henryk Niewodnicza\'nski Institute of Nuclear Physics, Krak\'ow, Poland}
{The scalar-isoscalar, scalar-isotensor and vector-isovector $\pi-\pi$
partial wave amplitudes are analyzed. Preliminary results indicate that
only the scalar-isoscalar amplitude fitted to the "down-flat" data satisfies
Roy's equations and consequently crossing symmetry.\\ {\textit {Talk given at
MENU2001, George Washington University, July 26-31, 2001.}}}

\section{Motivations and theoretical constraints}

Pions are produced in many reactions. A good experimental and theoretical
information on low, medium and high-energy pion-pion interactions should
give us a better understanding of non-perturbative QCD and of the chiral
perturbation theory. It allows a better insight into $q \bar q$ vacuum
condensate and then into the mechanism of spontaneous breaking of chiral
symmetry. It should also give a better knowledge of the meson spectrum in
particular of the $\sigma$-meson and glueballs.

Recently Kami\'nski et al.~\cite{Kaminski97}, using not only pion-exchange
but also $a_{1}$-exchange, have reanalyzed the data obtained in the
seventies on the $\pi^- p \rightarrow \pi^+ \pi^- n$ reaction at 17.2 GeV/c
without and with a polarized target. Essentially, for $m_{\pi\pi}$ between
600 and 980 MeV, two solutions ``up-flat'' and ``down-flat'' (hereafter
called \textit{up} and \textit{down}) were found for the pion-pion isoscalar
$S$-wave.

Besides unitarity and analyticity, crossing symmetry plays a very important
role in the $\pi\pi$ interactions as in each channel the same particles
interact. Projection on partial waves of the twice subtracted fixed-t
dispersion relations leads to Roy's equations for the scattering
amplitude of isospin I, viz.
$$
    Ref_{\ell}^{I}(s) = 
    \left(
    \begin{array}{c}
       \hspace{-0.2cm} a_{0}^0  \\
      \hspace{-0.2cm}  0  \\
      \hspace{-0.2cm}  a_{0}^2
    \end{array}
    \hspace{-0.2cm} \right)
    \delta_{\ell 0}+(2a_{0}^0-5a_{0}^2)\ 
    \frac{s-4m_{\pi}^2}{12m_{\pi}^2}
    \left(
    \begin{array}{c}
      \hspace{-0.2cm}  \delta_{\ell 0}  \\
      \hspace{-0.2cm}  \delta_{\ell 1}/6  \\
      \hspace{-0.2cm}  -\delta_{\ell 0}/2
    \end{array}
    \hspace{-0.2cm} \right) \\
     +  \sum\limits_{I'=0}^{2}\sum\limits_{\ell'=0}^{1}
      \hspace{0.15cm}-\hspace{-0.65cm}\int\limits_{4m_{\pi}^2}^{110 m_{\pi}^2}
    \hspace{-0.3cm} K_{\ell I}^{\ell^\prime I^\prime}(s,s') 
    \mbox{Im }f_{\ell'}^{I^\prime}
     (s') ds' + d_{\ell}^{I}(s).
$$
These equations are valid for $4m_{\pi}^2\le s\le 68m_{\pi}^2\ (=1.15 \mbox{
GeV})$ and express the real part of the scalar-isoscalar, scalar-isovector
and vector-isovector partial waves as integrals on their imaginary parts.
The two subtractions are expressed in terms of the scattering lengths
$a_{\ell}^I$: $a_{0}^0$ and $a_{0}^2$. The kernels K are known singular
functions. The driving terms $d_{\ell}^I(s)$ contain the contributions of
the partial waves $\ell' \ge 2$ as well as the high-energy
contributions~\cite{Roy71, Ananthanarayan00}. For the driving terms we use
the parameterization of Basdevant et al.~\cite{Basdevant74}.

Pions are the quasi-Goldstone bosons of the chiral symmetry of strong
interaction, so at low $m_{\pi \pi}$ we use constraints from chiral
perturbation theory. For instance the two-loop calculations of the $\pi\pi$
amplitudes using $K_{\ell 4}$ decay constraints lead to: $a_{0}^0=0.219\pm
0.005 m_{\pi}^{-1}$, $a_{0}^2=-0.042\pm 0.001m_{\pi}^{-1} $ and to the slope
parameters of the phase shifts; $b_{0}^0 =0.279\pm 0.011m_{\pi}^{-3}$,
$b_{0}^2=-0.076\pm 0.002m_{\pi}^{-3}$~\cite{Amoros00}.

\section{Applications and outlook}

For the scalar-isoscalar phase shifts $\delta_{0}^0$ we use the unitary
three-channel model of Kami\'nski et al.~\cite{Kaminski99} with chiral
symmetry constraints on $a_{0}^0,\ b_{0}^0$ and on $\pi\pi\to K\bar K$ and
$K\bar K\to K\bar K$ reactions at the $\pi \pi$ threshold as given by
Donoghue et al.~\cite{Donoghue90}. For $\delta_{0}^2$ we use chiral symmetry
constraints on $a_{0}^2$ and $b_{0}^2$ and do a fit to Hoogland A
data~\cite{Hoogland77} using the Pad\'e approximant parameterization of
Schenk~\cite{Schenk91}. For $\delta_{1}^1$ we use the Schenk
parameterization with $a_{1}^1=0.035m_{\pi}^{-1}$~\cite{Chell93}.

First we built up a three-channel fit to the solution \textit{down} and to
Roy's equations. The quality of fit to Roy's equations is judged by a
comparison between the exact real part of the partial waves calculated from
phase shifts and inelasticities (called input) and the output calculated
from Roy's equations. These equations are well satisfied by this fit called
``best-down'' fit. The input is close to the output with some small
deviations above 900 MeV, in particular for the scalar-isotensor wave.

Let us try to solve the \textit{up}-\textit{down} ambiguity. The solution
\textit{up} and \textit{down} differ mainly for $800\le m_{\pi\pi}\le 980$
MeV. We use a Pad\'e approximant with 8 parameters to fit $\tan
\delta_{0}^0$ of the solutions \textit{up} and \textit{down}. The parameters
are determined: i) to reproduce the "chiral" values $a_{0}^0,\ b_{0}^0$ and
the $\delta_{0}^0$ of the ``best-down'' fit at 500 and 600 MeV, ii) to have a
smooth junction to the ``best-down'' fit close to 970 MeV and iii) to obtain a
best fit of the data between 680 and 950 MeV.

\begin{figure}[t]
\parbox{.4\textwidth}{\epsfig{file=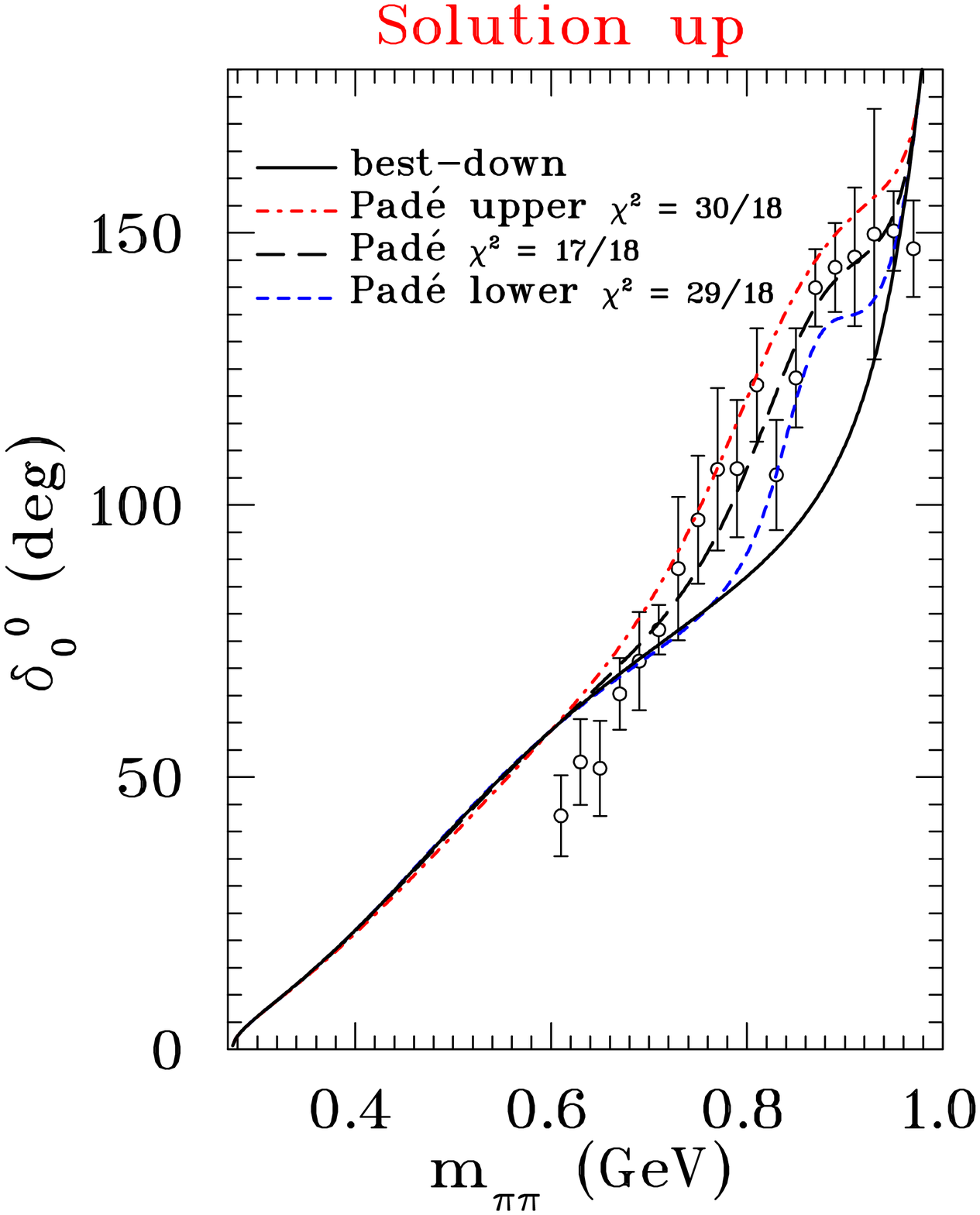,
width=.41\textwidth,silent=,clip=}}
\hspace*{.5in}
\parbox{.4\textwidth}{\epsfig{file=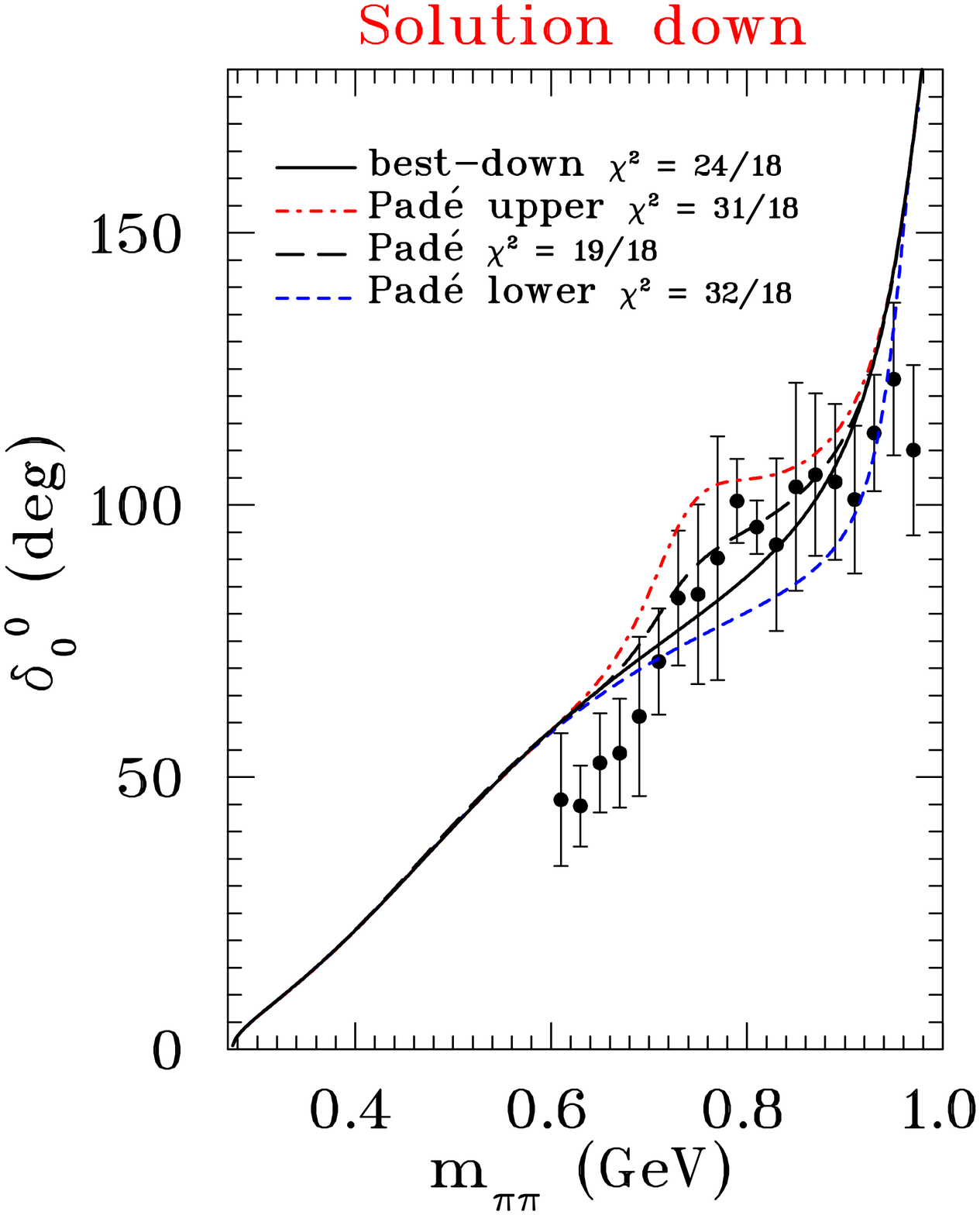,
width=.41\textwidth,silent=,clip=}}
\caption{\label{fig1} Left: fits to the solution \textit{up}.
Right: fits to the solution \textit{down}.}
\end{figure}

Results of different fits with their corresponding $\chi^2$ are shown in
Fig. 1. The ``Pad\'e lower'' and ``Pad\'e upper'' denote the fits to the
data shifted downwards and upwards according to their error bars. In Fig. 2
we check how this ``upper'' and ``lower'' fits satisfy Roy's equations.  For
the $S$-wave and for $840\le m_{\pi\pi} \le 950$ MeV (left panel) there is
no overlap between the ``upper'' and ``lower'' input and output bands. Here
the solution \textit{up} does not satisfy Roy equation, i.e. it is not
compatible with crossing symmetry. On the contrary for the solution
\textit{down} (right panel), for $m_{\pi\pi} \le 950$ MeV, both bands
overlap. The other waves satisfy relatively well Roy's equations. The
solution \textit{down} is compatible with crossing symmetry.

\begin{figure}[t]
\parbox{.4\textwidth}{\epsfig{file=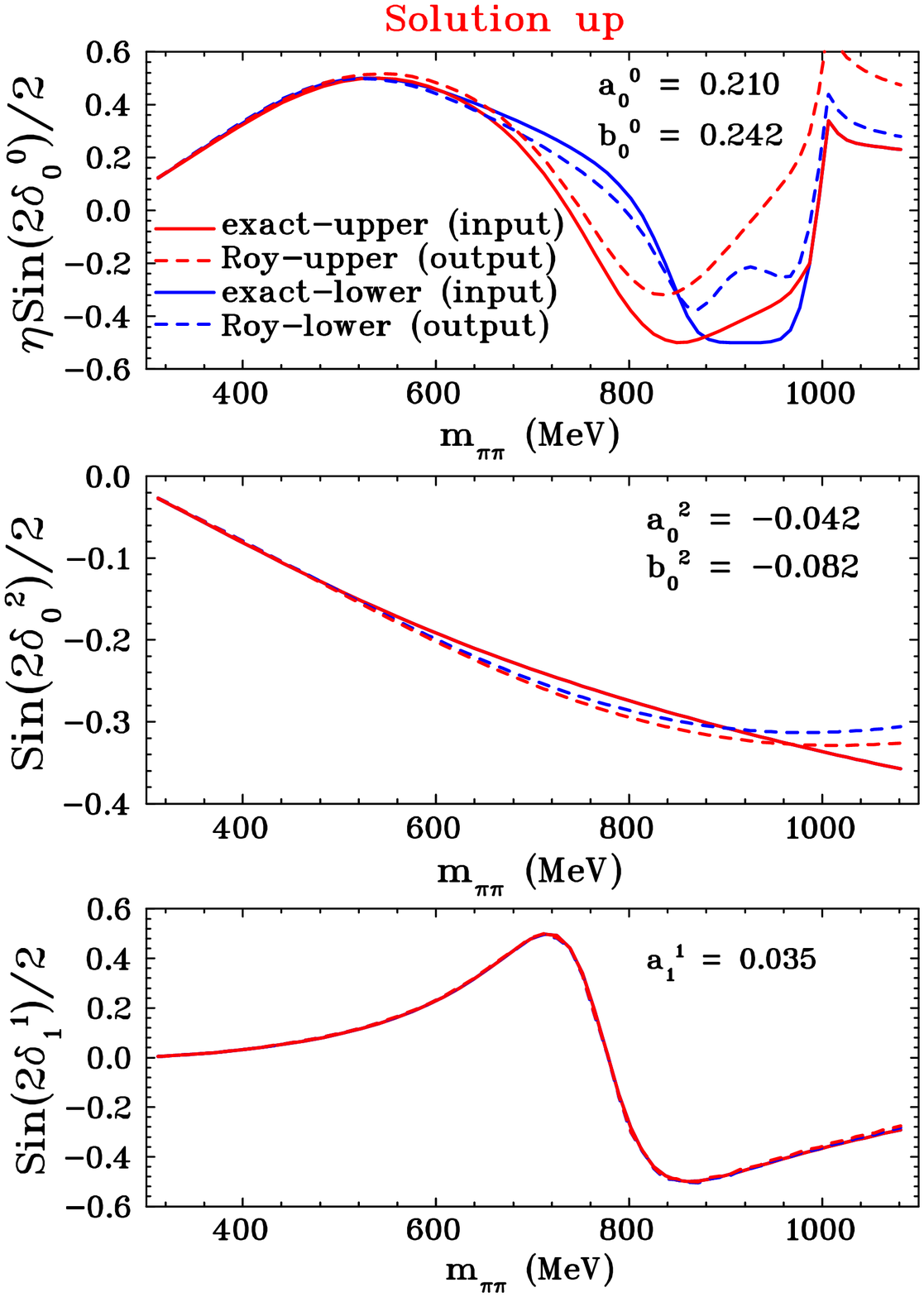,
width=.45\textwidth,silent=,clip=}}
\hspace*{.5in}
\parbox{.4\textwidth}{\epsfig{file=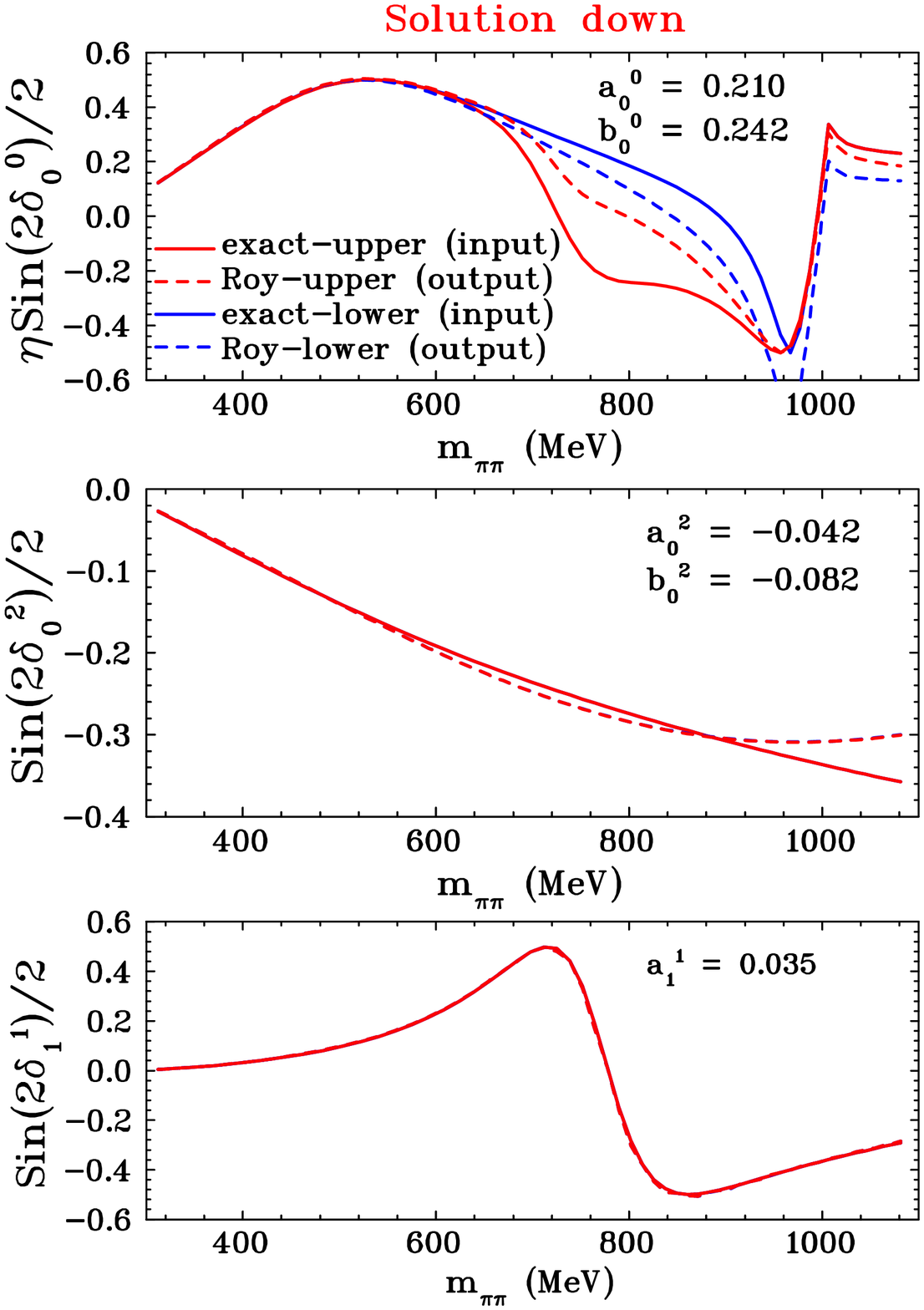,
width=.45\textwidth,silent=,clip=}}
\caption{\label{fig2}  Tests of  Roy's equations
for solutions \textit{up} (left panel) and \textit{down} (right panel).}
\end{figure}

Let us remark here that recent joint analysis by Kami\'nski et
al.~\cite{Kaminski01} of the CERN-Munich, the CERN-Cracow-Munich
$\pi^+\pi^-$ data and of the $\pi^0\pi^0$ data of the Brookhaven E852
Collaboration at 18.3 GeV/c~\cite{Gunter00} eliminates the ``up-flat''
solution and leads to a solution compatible with solution \textit{down}. Our
study here shows that only the solution \textit{down} satisfies crossing
symmetry.

Currently we are studying the sensitivity of the driving terms to the
parameterization of the resonant $f_{2}(1270)$ and $\rho_{3}(1690)$
amplitudes and to the high energy Regge contributions. We shall compare our
results to those of Ananthanarayan et al.~\cite{Ananthanarayan00}.

The present study shows that we can construct a three-channel $\pi\pi, \
K\bar K$ and effective $(2\pi)(2\pi)$ model which fulfills unitarity,
crossing symmetry and chiral symmetry constraints. This will give more
confidence in the parameters of the scalar-isoscalar mesons predicted by the
model. This will also allow to check the amplitudes below the $K \bar K$
threshold. Let us finally mention that any new precise data on $\pi\pi$ are
welcome.

\acknowledgments{The LPNHE is a Unit\'e de Recherche des Universit\'es
Paris 6 et Paris 7, associ\'ee au CNRS. This work was done in the
framework of the IN2P3-Polish laboratories Convention (project No. 99-97).}

\end{document}